\documentstyle[prd,aps,floats]{revtex}

\begin{document}

\input epsf

\def\be{\begin{equation}}
\def\ee{\end{equation}}
\def\ben{\begin{eqnarray}}
\def\een{\end{eqnarray}}


\def\negenspace{\kern-1.1em} 
\def\quer{\negenspace\nearrow}

\def \Behauptung{\buildrel {{\scriptstyle ! } } \over { = } }   


\newcount\secno
\secno=0
\newcount\susecno
\newcount\fmno\def\z{\global\advance\fmno by 1 \the\secno.
                       \the\susecno.\the\fmno}
\def\section#1{\global\advance\secno by 1
                \susecno=0 \fmno=0
                \centerline{\bf \the\secno. #1}\par}
\def\subsection#1{\medbreak\global\advance\susecno by 1
                  \fmno=0
       \noindent{\the\secno.\the\susecno. {\it #1}}\noindent}


\def\sqr#1#2{{\vcenter{\hrule height.#2pt\hbox{\vrule width.#2pt
height#1pt \kern#1pt \vrule width.#2pt}\hrule height.#2pt}}}
\def\square{\mathchoice\sqr64\sqr64\sqr{4.2}3\sqr{3.0}3}
\def\hatsquare { {\hat\sqcap\!\!\!\!\sqcup} }


\newcount\refno
\refno=1
\def\y{\the\refno}
\def\myfoot#1{\footnote{$^{(\y)}$}{#1}
                 \advance\refno by 1}


\def\newref{\vskip 1pc 
            \hangindent=2pc
            \hangafter=1
            \noindent}

\def\neq{\hbox{$\,$=\kern-6.5pt /$\,$}}


\def\asteq{\buildrel \ast \over =}


\font\fbg=cmmib10\def\clom{\hbox{\fbg\char33}}


\newcount\secno
\newcount\fmno\def\z{\global\advance\fmno by 1 \the\secno.
                       \the\fmno}
\def\sectio#1{\medbreak\global\advance\secno by 1
                  \fmno=0
       \noindent{\the\secno. {\it #1}}\noindent}



\centerline{\bf GRAVITATIONAL STABILITY OF BOSON STARS}  
\bigskip
\bigskip\bigskip  
\centerline{by}  
\bigskip
\centerline{Fjodor V. Kusmartsev$^{* + \$ }$}
\bigskip
\centerline{Eckehard W. Mielke$^{* \diamond }$}
\bigskip
\centerline{Franz E. Schunck$^{* \triangle }$} 
\bigskip\bigskip

\noindent $^{*})$ Institute for Theoretical Physics, University of 
Cologne, D-5000 K\"oln 41, Federal Republic of Germany

\noindent $^{+})$ L. D. Landau Institute for Theoretical Physics, 
Moscow 117334, USSR

\bigskip\bigskip
\bigskip
\centerline{\bf Abstract}
\par
We investigate the stability of general--relativistic 
boson stars by classifying singularities of differential mappings
and compare it with the results of perturbation theory.
Depending on the particle number, the star has
the following regimes of behavior: stable, metastable,
pulsation, and collapse.


\vskip 3.cm

$^{\$ })$ Supported by the Alexander von Humboldt--Foundation, Bonn.

$^{\diamond })$ Supported by the Deutsche Forschungsgemeinschaft,
project He $528/12-1$.

$^{\triangle })$ Supported by the ``Minister f\"ur Wissenschaft
und Forschung" of Nordrhein--Westfalen.

\vfill\eject

\bigskip

\section {\bf Introduction}
\par
Presently, there is much interest in the problem of 
stability of matter confined by its self--generated
gravity. This self--consistent approach dates back to the {\it geons}
of Wheeler $[1]$. Recently, the work of Lee et al. $[2, 3]$
stimulated further progress.
They pointed out that a star, regarded as a gravitational 
soliton $[4-6]$, can have a mass which is 
larger than the Chandrasekhar type limit for gravitational 
collapse  $[7, 8, 9]$. This opens up a new avenue for studying 
the structure of a star under unusual matter conditions.
Its stability is the most important question.
So far, the dynamical stability of boson stars has been analysed 
$[10-12, 13]$ by means of perturbation theory. 
In this paper, we will apply a method 
which was proposed by one of us ($[14]$ and references therein)
for nongravitational solitons.
In general, the method 
consists of investigating the critical points of a mapping
and the construction of bifurcation diagrams.
In our case, a two--dimensional subspace of the dynamical variables
of the boson field 
is mapped into the space of the integrals of motion, such as
the gravitational mass $M$ and the total particle number $N$. 
Using Arnold's classification $[15]$ of singularities
of differential maps (catastrophe theory), we are able to derive 
general criteria for the stability of the star.

\bigskip

\section {\bf Coupled Einstein--scalar field equation}
\par
As a general--relativistic model of a boson star, we consider a
self--interacting scalar field $\Phi $ describing a state
with zero temperature. This field is self--consistently coupled
to its own gravitational field via the Lagrangian
\be {\cal L}={1\over 2\kappa }\sqrt{\mid g\mid }\; R
                  +{1\over 2} \sqrt{\mid g\mid}\;
   \left [ g^{\mu \nu}(\partial_\mu \Phi^\ast )(\partial_\nu \Phi )
             -U(\mid \Phi \mid^2) \right ]\; ,  \ee
where $\kappa =8\pi G$ is the gravitational constant in natural units,
$g$ the determinant of the metric $g_{\mu \nu }$,
$\mu ,\nu =(0,1,2,3)$, $R$ the curvature scalar,
and $U(\mid \Phi\mid ^2)$ the 
self--interaction potential. We will investigate to what
extend the form of $U$ influences the stability of the star. 

\par
From the principle of least action we obtain the {\it coupled}
Einstein--Klein--Gordon equations:
\ben
               R_{\mu \nu } -{1\over 2}g_{\mu \nu }R 
             &=-\kappa T_{\mu \nu }(\Phi ),&  \\
      \left (\square +{dU\over d\mid \Phi \mid^2}\right )\Phi 
             &=0,& 
\een
where
$ T_{\mu \nu }(\Phi )
   =(\partial_\mu \Phi^\ast )(\partial_\nu \Phi )
  -( g_{\mu \nu }/ \sqrt {\mid g\mid } ) {\cal L} (\Phi ) $
is the energy--momentum tensor and
$ \square = (1/ \sqrt {\mid g\mid } ) \partial_\mu 
\left (\sqrt{\mid g\mid }g^{\mu \nu } \partial_\nu \right ) $
the generally covariant d'Alembertian.

\par
In this paper, we restrict ourselves to the static, spherical symmetric 
metric
\be ds^2=e^{\nu (r)}dt^2-e^{\lambda (r)}dr^2
        -r^2(d\theta ^2+\sin ^2\theta d\phi ^2) ,  \ee
in which the functions $\nu =\nu (r)$ and 
$\lambda =\lambda (r)$ depend on the Schwarzschild type radial
coordinate $r$.
For the boson field, we make the stationarity ansatz
\be \Phi (r,t)=P(r) e^{-i\omega t} \; ,  \ee
which describes a spherically symmetric bound 
state with frequency $\omega $.
The resulting coupled system reads
\ben
\nu '+\lambda ' 
           &=\kappa (\rho +p_r)re^\lambda \quad , \quad &  \\
             \lambda '
           &=\kappa \rho re^\lambda-{1\over r}e^\lambda
                        +{1\over r} \quad , \quad &  \\
P''+\left ({1\over 2}(\nu '-\lambda ')+ {2\over r}\right )\; P' 
           &=e^\lambda {dU\over dP^2}P 
    -e^{\lambda -\nu}\omega^2\; P \quad . \quad & 
\een
The energy--momentum tensor becomes diagonal, i.e.
$T_\mu ^{\; \nu }=diag \; (\rho , -p_r, -p_\bot, -p_\bot )$ with
\ben
   \rho &= {1\over 2} (\omega^2 P^2 e^{-\nu}
                  +P'^2 e^{-\lambda} +U )\; , & \\
    p_r &=  \rho -  U \; , \;  
 p_\bot  =  p_r -  P'^2 e^{-\lambda } \; . & 
\een
The form of $T_\mu ^{\; \nu }$ is familiar from an ideal fluid,
except that the radial and tangential pressure generated by the
scalar field are in general different, i.e. $p_r \neq p_\bot $.
This {\it fractional anisotropy} $a_f :=(p_r - p_\bot )/p_r $
has already been noted 
by Ruffini and Bonazzola $[5]$. Moreover, Gleiser [13] found that all
boson stars have the same amount of anisotropy at the radius
of the star.

\par
Because of the contracted Bianchi identity 
$ \nabla^\mu \left (R_{\mu \nu }
        -{1\over 2} g_{\mu \nu } R\right ) \equiv 0  $,
a further equation involving 
$T_\Theta^{\; \Theta}=T_\phi^{\; \phi}=-p_\bot $
is identically satisfied.
Eq. (2.7) possesses a Schwarz\-schild--type solution
\be  e^{-\lambda (r)}=1-{\kappa \alpha (r)\over r} , \qquad 
 \alpha (r):=\int\limits_0^r \rho x^2dx . \ee
where $\alpha (r)$ is the mass function.
For the polynomial self--interaction
\be U:=m^2{\mid \Phi \mid}^2 +{1\over 2}\alpha \mid \Phi \mid^4
      +{1\over 3}\beta \mid \Phi \mid^6    \; ,
\ee
these equations has been solved numerically for nonsingular, 
finite mass and zero--node solution $[3,4,5,16,17]$.
Two and higher node solutions occurred already in Ref. $[6]$.
For a massless scalar field with $U=0$, an exact solution is
known, cf. Ref. [18]. For a massless real scalar field,
Christodoulou [19] could show that a spherically symmetric
time--dependent field configuration must either disperse to
infinity or, for non--vanishing Bondi mass, forms a black hole.

\bigskip

\section {\bf Integrals of motion}
\par
The concept of an energy--momentum $4$--vector for a field 
configuration is a notoriously subtle
[20] in general relativity. However, the exponential decrease of the 
radial function $P(r) \simeq exp[-\sqrt{m^2-\omega^2}\, r] $
for $\mid \omega \mid < m$ yields an 
{\it isolated}, static system for which the
{\it Tolman mass formula}
\be M:=\int (2T_0^{\; 0}-T_\mu^{\; \mu }) 
         \sqrt{\mid g\mid} \; d^3x  
       = 4\pi \int \limits_0^ \infty
         \bigl [ 2 \omega^2 P^2 e^{-\nu} -U \bigr ]
           e^{(\nu +\lambda )/2} r^2 dr 
\ee
applies (cf. [6]). It can be derived from the {\it local} conservation 
law $ \partial_\nu ({\cal T}_\mu ^{\; \nu }+\tau_\mu ^{\; \nu })=0$,
where ${\cal T}_\mu ^{\; \nu }=\sqrt{g}\; T_\mu ^{\; \nu }$ and
$\tau_\mu ^{\; \nu }$ is the gravitational energy--momentum complex.
For a boson star, the explicit expression (13)
does not involve derivatives, in contrast to the Schwarzschild
mass
$ M_{Schwarzschild}:=4\pi \alpha (\infty ) $
that is commonly studied $[11]$ in this context.
Friedberg et al. implicitly rederived the equivalence of the
Tolman and the Schwarzschild mass (see (2.27) of Ref. $[3a]$;
cf. also $[7]$).

\par
A second ``integral of motion" arises from the fact that the 
Lagrangian (1) is invariant under the global phase transformation 
$\Phi \rightarrow \Phi e^{-i\vartheta }$.
Therefore the Noether current density
\be j^\mu  
  ={i\over 2} \sqrt{\mid g\mid }\; g^{\mu \nu }
 \bigl [\Phi^\ast \partial_\nu \Phi -\Phi \partial_\nu \Phi^\ast \bigr ]
\ee
is {\it locally} conserved, i.e. $\partial_\mu j^\mu =0$.
The time--component $j^0$ integrated over space yields the
{\it particle number} $N$ or the {\it charge} $Q$:
\be N={Q\over e}=
  4\pi \omega \int\limits_0^\infty \;e^{(\lambda -\nu )/2}r^2P^2 dr
   \; . \ee

\par 
Since the current density (14) is a ``measure" for the radial 
distribution of the ``particles" in the boson star, 
its {\it effective radius} can be defined by
\be R:= {1\over N} \int r j^\mu d\Sigma_\mu
     = {4\pi \omega \over N} \int_0^\infty e^{(\lambda -\nu )/2}
        r^3 P^2 dr \; .
\ee
On account of the fractional anisotropy $a_f$ another interesting
radius $R_0$ could be obtained from the node $p_\bot (R_0)=0$ in
the tangential pressure $p_\bot $. This radius $R_0$ separates
the interior part of the boson star from a {\it marginal layer}
in which $p_\bot $ becomes negative before it decreases
exponentially [17].

\goodbreak

\bigskip

\section{\bf Smooth mapping (Whitney surface)}

\par
In order to investigate the stability of soliton--type solutions
against radial perturbations, we consider the two--dimensional mapping
\be F \; : \; (k,\omega ) \;  \mapsto \;  (M,N) \; , \ee
where $k$ is a variational parameter 
which dilatates the radius $R$ of the star
and $\omega $ the frequency eigenvalue. 
The parameter $k$ induces a scaling of the metric, the frequency,
and the scalar field in accordance with their normal physical
dimensions
\be ds^2 \; \rightarrow \; k^2ds^2 \; ,\;
   \omega \; \rightarrow \; \omega /k  \; , 
 \;  P(r) \; \rightarrow \; kP(kr) \; ,
\ee
such that the particle number $N$ is kept fixed.

\par
In order to classify the singularities of this mapping $F$, let us
consider the Jacobi matrix
$$ J=\pmatrix{ \partial M / \partial k  &
               \partial M / \partial \omega   \cr
               \partial N / \partial k  &
               \partial N / \partial \omega   \cr } .
    \eqno(\z) $$
According to Whitney's theorem $[14,15]$,
the singularities of the mapping $F$ can be one of three types, 
depending on the rank $R_J=2,1$ and $0$, respectively.
Since we require the soliton solution to be an extremal point of the
Lagrange manifold, we have:
$$ {\partial M \over \partial k}=0 \quad and \quad 
   {\partial N \over \partial k}=0 .   \eqno(\z) $$
{\it For the soliton} the rank of $J$ is
$ R_J < 2 \; $ and,
consequently, the singularities of the mapping $F$
may have either $R_J=1$ or $R_J=0$.
In that case, our soliton 
solution corresponds to the extremal or critical 
points of the Whitney surfaces which
has a very definite form
(see $[14,15]$). In numerical examples, the dependence
$$ M = M(\omega ) \; , \; N = N(\omega ) \; , \quad with \quad 
   \omega =\omega (\sigma (0)) , \eqno(\z) $$
on the frequency $\omega $ can be smoothly converted into a function
of the central density $\sigma (0)=\sqrt {\kappa }\, P(0)$
such that the critical points coincide (Fig. 1). 
If the rank of $J$ is zero,
the critical points are degenerate. The maxima and minima of
$M=M(\sigma (0))$ and $N=N(\sigma (0))$, see Fig. 1,
correspond to the $A_2$ singularity, in the notation
of Arnold. Other points of the curves in Fig. 1 correspond to
the critical points $A_1$.

\bigskip

\section{\bf Bifurcation diagram}

\par
In order to classify the nondegenerate $A_1$, we need to
consider the bifurcation diagram $M=M(N)$ [3]. Equivalently, we may
consider the {\it binding energy}
$$ B=M-mN=B(N)  \eqno(\z) $$
as a function of the particle number (Fig. 2). A further
``magnification" of $B(N)$ is achieved in Fig. 3.
According to the Whitney theorem, the cuspoidal points
of these diagrams classify the $A_2$
singularity, whereas the other points of the diagram correspond
to the $A_1$ singularity. Each cusp represents some
{\it Whitney surface}, which is a part of the mass--energy surface.
As shown in Ref. $[14]$, the minimum on 
this surface corresponds to the stable soliton, the maximum 
corresponds to the unstable soliton. At the cuspoidal point, 
the minimum coalesces with the maximum, and
the soliton loses its stability.
Thus, the lower branch of the lowest cusp corresponds 
to the {\it absolutely stable soliton}. The upper branch of the 
first cusp, which is, at the same time, the lower branch 
of the second cusp, corresponds to the unstable soliton. The upper 
branch of the second cusp also corresponds to unstable 
solitons, which, however, suffer from a different kind of
instability than the soliton of the lower branch of the second cusp.

\par
For the boson star, the degrees of freedom of the configuration
space are very large. The fact that the lower branch of the
lower cusp corresponds to absolutely stable 
soliton means that there we have minima for all directions in the 
configuration space. 

\par
The higher branch of the first lower cusp corresponds to a
maximum. This maximum occurs in that section of the mass--energy surface
which depends on the radius $R$ of the star.
For the second cusp the appearance of a new instability depends 
on the mutual branching to other cusps. There
are the two possibilities that the next (third) branch
goes higher or lower than the second one.
In the first case --- according to Whitney's theorem ---
the minimum transforms into a maximum at the transition
from the second branch to the third one,
and new instability appears. Vice versa, in the second case,
one maximum transforms into minimum after the transition through the
cuspoidal point and one instability disappears. 
The numerical data (Fig. 3) show that for the third 
cuspoidal point one instability disappears, in
accordance with the general picture developed here.
This disappearence of one instability has not been pointed
out in previous works on the stability of boson stars.

\par
In the literature $[3, 21]$ the corresponding diagrams are obtained
by applying the method of small perturbations to the problem of
stability. In our approach, the bifurcation diagram is a key point
of the analysis and is gained by analyzing the topology of the
Whitney surface. In the first step of the analysis, we calculate the 
bifurcation diagram for the mapping which relates the 
integrals of motion $M$ and $N$ to some degrees of freedom of the
boson star. Following $[14]$ we connect this bifurcation
diagram to the mass--energy surface $M=M(R,N)$.

\par
The mass--energy surface described by the bifurcation diagram
corresponds to some complicated manifold of catastrophe, which
probably has never been seen before in the theory of singularities.
The order of this grand catastrophe depends on the number
of cusps in the bifurcation diagram.
The {\it static description} of the 
bifurcation diagram is that at every cusp there is a transition 
from minimum to maximum or vice versa. For every cusp there occurs
one Whitney surface. As an application to our case, we find at
the first cusp that a minimum transforms
to a maximum. This maximum will not be affected at the 
following cusps, if at the second cusp the following branch goes 
to higher mass values. If this branch reached lower mass values,
we will get a more complicated ``Whitney surface''
in analogy to the case of the swallow's tail [22].

\par
For neutron stars or white dwarfs, we get a complicated
bifurcation diagram consisting of the many 
cusps (see Fig. 8 of Ref. [8]). Again, each cusp corresponds to
a Whitney surface. The first four lower branches of this
bifurcation diagram (in the direction of increasing density)
describe a section of manifold of catastrophe
which is usually called a butterfly.
Consequently, for these four branches this two--dimensional 
manifold of catastrophe corresponds to the two--dimensional Whitney 
surface in the bosonic case.

\par
For the second cusp of the bosonic bifurcation diagram there is
a Whitney surface also, i.e. at a cuspoidal point of this cusp there
is a transition from minimum to maximum. At the following 
cusps, if there is a transition to higher mass values, 
the maxima stay maxima.
We understand such applications of catastrophe theory not 
only for the boson star, but also for fermion Q--stars,
white dwarfs, and neutron stars. The latter have been
considered by Harrison, et al. [8]. They `saw' the Whitney surface
without drawing knowledge from catastrophe theory.

\par
The simplest way to identify the instabilities, which have been 
qualitatively predicted by catastrohpe theory, is to consider 
perturbation theory. Because the perturbative equations set up a
Sturm--Liouville eigenvalue problem, the characteristic frequencies 
of the perturbation series have increasing absolute values.
This holds also in the case of the boson star [11, 12, 13].
Moreover, it was shown for the boson star [11] and for the 
neutron star [8] that, at each cusp, one of these frequencies changes
sign. Thus, each such instability can be identified with
a corresponding one obtained from applying catastrophe theory.

\par
Until now we have considered only the mapping of the two--dimensional
space $(k, \omega )$ into two--dimensional space $(M, N)$. Since
the first space counts the number of the degrees of freedom of
the star, we may extend the domain $(k, \omega )$ of the
mapping by including
the characteristic frequencies. In the next step, new states
of the star with different dynamical behavior could be taken into 
account than those we see already in the bifurcation diagram.
These states correspond to other points on the mass--energy
surface. For example, for burning stars oscillations with large
amplitudes occur at their finite stage as red giants. The evolution
of these oscillations cannot be described by perturbation theory.
This kind of dynamical behavior of the star will be discussed
in Section 6.

\par
In flat spacetime, the dependence $M$ on $N$ has been investigated 
by Friedberg  et al. $[21]$. Although they pointed out that the
minimal energy branch of $M$ versus $N$ is stable, which is true,
they have not investigated the stability of these solitons
for all values $M$ and $N$.

\bigskip

\section{\bf The different regimes of the star's behavior}

\par
In our method, like in the theory of singularities of smooth 
mappings, the bifurcation diagram plays an important role. It is a 
skeleton of the catastrophe or skeleton of the mass--energy 
surface. The extremal point of this surface corresponds to  the 
soliton solution. There exist also other types of solutions, with a 
different dynamical behavior. Perturbation theory gives
small oscillations near the soliton solution. We went beyond
perturbation theory, which helped us to investigate
the stability of the star.

\par
Using the catastrophe theory, we can construct the mass--energy 
surface. Each section of this mass--energy surface contains the
different degrees of freedom of the star, which can be identified,
by perturbation theory, with the characteristic frequencies.
Using such sections, one can predict dynamical regimes of the 
star which cannot be described within the framework of perturbation 
theory.

\par
In order to predict the different regimes of behavior of the star,
we construct the section $M(1/R)$ (where $R$ is the effective
radius of the star) of the mass--energy surface
$M(R, N)$ at fixed $N$. In other words, we construct an adiabatic
potential [23]. The shape of $M(1/R)$ follows from the bifurcation 
diagram.

\par
The type of regimes depends on the critical values
$N_{C_1},N_{C_2},N_{C_3}$, of the particle number.
For the coupling constants $\tilde \alpha =(\alpha /\kappa m^2)=10$
and $\beta =0$ in the self--interacting potential (2.12), we find
that $N_{C_1}=0.54$, $N_{C_2}=0.68$,
$N_{C_3}=1.20$ are the cuspoidal points, see Fig. 2.

\bigskip

\subsection{\bf Stable soliton and oscillation}

\par
From the bifurcation diagram (Fig. 2) it can be inferred that for 
$N<N_{C_1}$ the function $M(1/R)$ has only one 
minimum, which corresponds to the stable soliton solution
(lower branch of the first cusp).

\par
The dependence of $M$ on $R$ at some fixed value 
$N<N_{C_1}$ is schematically presented in Fig. 4.
The smooth extremal point which corresponds to the minimum of the
curve $M(1/R)$ is associated with the soliton solution. The value
of $M$ in this point we call $M_{soliton}$. The marginal extremal
point at $R = \infty $ corresponds to the
``homogeneous state", or plane wave solution in flat spacetime.
In the self--generated gravitational field, it can be defined as an
{\it effectively free boson field solution} for which the
binding energy $B=M-mN$ is vanishing.
At infinite value of the radius $R$ of the star the values $M$
and $N$ are final ones. This follows from the fact that
in the limit $R \rightarrow \infty $ the density of the star
goes to zero (see Ref. [14], p. 29).
The exact construction of the corresponding solution will be
deferred to a future publication. These two extremal
solution characterize a static configuration of the star.  As we 
see from this figure, the ``free boson'' field is unstable. This 
extremum corresponds to the maximum. It means that the 
homogeneous state of the star will collapse from the size 
$R=R_0=\infty$ to the size $R=R_1$ (see Fig. 4).
From $R_1$, the size of 
the star will again increase. The value of $R$ increases
up to the initial value $R_0$.
Thus the star will be in an oscillating 
regime. For such oscillations of the star, the curve presented in 
Fig. 4 is a kind of ``adiabatic'' potential which can also be obtained 
from a scale transformation. There are also other 
oscillations of the star in this potential that correspond to
other values of $M$.  For example,  the other oscillation regime of 
the star corresponds to the horizontal line 1. In this case the
amplitude of the oscillation is lower than the oscillations
in the regime associated with the unstable free boson field described
above. On the other hand, the oscillating regime, corresponding
to the line 2, has a smaller amplitude than the one corresponding
to the line 1 and so on. The limiting case is a stable soliton without
oscillation. In the region of $N<N_{C_1}$,
Fig. 4 gives a complete picture of the star's behavior.

\par
The interesting point here is that any arbitrary
configuration of fixed particle number $N$ cannot have a mass
smaller than $M_{soliton}$. The reason is that
the configurations of the star are limited by the mapping $F$.
For instance, the mapping
$F:(k,\omega ) \rightarrow (M,N)$ allows only a class of configurations
away from the static soliton which preserve the integrals
of motion $M$ and $N$. Of course, the perturbed configurations
of the stable soliton will evolve along the line
dictated by the differential map.
But there may arise other ways of choosing the parameters $k$
and $\omega $. Moreover, one could think of extending the domain
$(k,\omega )$ of the mapping $F$ to a higher--dimensional space 
$(k_1,\ldots ,k_N,\omega_1,\ldots ,\omega_N )$, where
$k_1,\ldots ,k_N,\omega_1,\ldots ,\omega_N$ correspond to
additional degrees of freedom of the star. Such a space provides
us with further lines of evolution of the perturbations, but
the stability will still be determine by the bifurcation diagram
$M(N)$. Such a diagram comes from a numerical solution
describing the stationary points of a Lagrange manifold of
Einstein's equation. In this way one can describe virtually
all perturbations which preserve the integrals of the motion $M$ and $N$.
If we considered configurations which cannot be categorized
by this choosen map, for example, when the total mass $M$ and the
total number of particles $N$ are not necessarily fixed, then
one could presume that the star, via some oscillation
process, settles down to a stationary stable configuration
with some reduced mass $M_{reduced}$.

\par
The oscillations arise for $M>M_{soliton}$ of the mass of the star. 
In this region $N<N_{C_1}$, at some value of $M>M_{soliton}$,
there may occur a gravitational collapse.
We expect that the configuration will oscillate rather than collapse to
a black hole not up to arbitrarily large mass $M$. There should exist
the critical value of $M_c = M_{Schwarzschild}$ which
depends on the radius of the star $R_c$ with given mass
$M_c(R_c)$. At this radius $R_c$ the star will stop
oscillating and start collapsing to a black hole.
However, in order to show this rigorously, the analysis of
Christodoulou [19] has to be extended to the case of the massive
or even nonlinear scalar field.

\bigskip

\subsection{\bf Collapse}

\par
On the other hand, for $N>N_{C_3}$, the
section $M(1/R)$ following from the bifurcation diagram of the 
mass--energy surface, has only the marginal extremum corresponding
to the effectively free boson field solution.
As we can infer from Fig. 5, the section $M(1/R)$ is a montonically 
decreasing function. The 
homogeneous free field solution corresponds to the marginal 
extremum. It is a maximum of $M(1/R)$.
For the increase of the kinetic energy 
of the star, the radius decreases. 
This means that in ``this region" of $N$
there exists a collapse for any value of the mass.
In the first stage, when $R_{star}>>(\kappa /4\pi )M_{Schwarzschild}$,
it is a wave collapse (see [24] for details and references therein)
which later induces the gravitational collapse.

\goodbreak
\par
Due to this {\it collapse}, this state is unstable.
The different horizontal lines, drawn in Fig. 5, correspond to 
different collapse regimes of the star with different initial 
radii. A similiar case is known for the creation of the 
two--dimensional plasma cavitons [25].

\bigskip

\subsection{\bf Pulsation, oscillation, and collapse}

\par
In the region $N_{C_2}<N<N_{C_3}$ the dependence of $M$ on $R$ has as 
many extremal points as there exist branches of cusps at a given 
value of $N$. In the case where there are two cusp branches at given $N$,
the dependence $M(1/R)$ has two extremal points and
a marginal extremum corresponding to the ``free boson field'' solution 
$M=mN$ (Fig. 6).
The point of the minimum of $M(1/R)$ corresponds to the stable 
soliton with the mass $M_{soliton}$. The instability of the
homogeneous state results in an oscillation regime similar to that
which has been described in Sect. 6.1.  The maximum of $M(1/R)$ 
corresponds to the unstable soliton with mass $M_u$.
The instability of this soliton can occur in two possible ways.
One of them is collapse, decreasing the radius of the 
star. The {\it second} one is increasing the star's
radius (dispersion). Such an instability gives rise to an
oscillation in the same manner as the marginal free particle
extremum. There can be a lot of such oscillations (see Sect. 6.1)
before the boson star will collapse.
That means that this oscillating regime may be also 
unstable and, after several periods, collapses to
the state of unstable soliton. Thus the existence of this unstable
oscillating regime is due to the existence of the unstable
soliton [24, 25].

\par
Thus, in the most interesting region $N_{C_2} < N < N_{C_3}$,
the function $M(1/R)$ has two extremal points, which correspond
to the stable soliton (lower branch of the first cusp, see
Fig. 2) and the unstable soliton (higher branch of
the first cusp), respectively, and one marginal extremum (free
boson field solution). The shape of such a function indicates
that at each value of $N$ there exist the following
configurations: a stable soliton
with a mass $M_{soliton}(N)$ and unstable one with
a mass $M_u(N)$. At $M(N)>M_u(N)$ 
it is very difficult to predict the evolution of the star.

\par
In the region $N_{C_1}<N<N_{C_2}$ 
the dependence of $M(1/R)$ may have more than 
two smooth extrema which are minima and maxima. 
For definiteness, let us consider the case when 
$M(1/R)$ has three extremal points. Two of these points 
correspond to stable solitons with the mass $M_{s_1}$ and $M_{s_2}$
and one corresponds to the unstable soliton with mass $M_u$.
There are two types of oscillations, corresponding to 
the minima $M_{s_1}$ and $M_{s_2}$ which exist at $M<M_{u}$. 
In the first case, the maximal kinetic energy of the star
corresponds to the radius of the first stable 
soliton. It is an oscillation in the first minimum.
For the second case, the maximal kinetic energy
corresponds to the radius of the second stable soliton. 
It is an oscillation in the second minimum.

\par
There is also a very interesting regime, for which the star has a 
mass $M=M_{u}$. 
In this case, there will exist a  spatial  type  of 
oscillation  in which it is difficult to predict in which direction
(first or second minimum) the star will move from the point of 
maximum. It is connected with the indefiniteness 
of  the  behavior of the star in the state of the unstable  soliton 
(dispersion or collapse). Such a regime of the star's behavior 
is characterized by a {\it pulsation}. The pulsation consists of at least
two different types of oscillations.

\par
In the same manner one can analyze the case when $M(1/R)$ has  more 
than three extremal points.

\par
One general conclusion which can be drawn here is that in the region
$N_{C_1}<N<N_{C_2}$ there is a significant distribution 
of the instabilities of the star.
The number of instabilities depends on the number
of cusps and on the mutual branching of these
cusps. If the next branch of some cusp corresponds to the higher
values of the star's mass, then the number of instabilities
increases by one. Vice versa, if the next branch of 
this cusp corresponds to lower values of the star's mass, then
the number of instabilities decreases by one.

\bigskip

\section{\bf Discussion}

\par
Our stability criteria include the results of Refs. [10, 11, 12, 13]
obtained by perturbation analysis.

\par
Until now, there exist some more or less successful attempts to prove
the stability of the boson star. These results fight with the 
difficulty of the mathematical problem; they tried to solve the 
problem of stability in a quantitative way like Harrison et al.
$[8]$ or Shapiro and 
Teukolsky $[7, 9]$. Gleiser $[13]$ and Jetzer $[12]$ 
got a upper limit for stability for the linear case and an additional
$\mid \Phi \mid^4 $ potential. This limit was much higher than the first
maximum in the $(M,\sigma (0))$--diagram.

\par
Later, Gleiser and Watkins $[11]$ showed that in the linear case
a change in stability occurs at the first extremum.
At the following extrema, the higher modes (in the context of $[8, 9]$)
are negative so that the star becomes more unstable. This result 
perfectly complies with those which are known from the analysis
of neutron stars. 
Furthermore, Jetzer $[12]$ showed that the zero node solution of the
boson star is stable until one reaches the first cusp.
Lee and Pang $[10]$ did not require the particle number $N$
to be constant, so they found that these solutions are unstable.

\par
The picture of the star's behavior, obtained here on the basis of the 
application of catastrophe theory to solitons $[14]$, has a general
character and has an analogous form for neutron stars
[7; 8, see Fig. 8], fermion Q--balls [26], 
and for dilaton stars $[27]$.
In fact, for neutron stars the diagram $M(N)$, obtained from
numerical integration of a certain equation of state for cold,
catalyzed matter, exhibit similar bifurcations as in the case
of boson stars (see Fig. 8 of Ref. $[8]$ and Fig. 47 of Ref. $[7]$).
For the first cusp, Harrison et al. $[8]$ could even deduce correct 
stability criteria from the analysis of the mass--energy surface
$M(\rho , N)$, see Fig. 9 of Ref. $[8]$. Although this was done 
without knowledge of catastrophe theory or Arnold's classification 
of singularities, these 1965 results are in complete agreement 
with the more general criteria developed here.

\par
In comparison with Ref. [19] one should point out that in the
limit $m \rightarrow 0$ all solutions, corresponding to the stable
solitons with $0<M_{soliton}<mN$ disappear and only
unstable solutions remain. Such solutions disperse
either to infinity or collapse. The collapse configuration can form a
black hole. This agrees completely with the results of Ref. [19].

\par
After the submission of our paper, the recent paper of
Seidel and Suen [28] appeared, in which
the numerical evolution of various configurations of boson
stars using the full nonlinear Einstein equations have been studied.
Moreover, perturbations which include a redistribution of
scalar particles in the star and also accreation and annihilation
of the bosons are considered so that the total mass $M$ and the total
number of particles $N$ are not necessarily fixed. Their result,
that the $U$--branch star (unstable soliton) will either collapse to form
a black hole or will disperse, agrees with our conclusion.
The exception is that the unstable soliton will eventually settle down
to a stable soliton. This is due to the possibility that
the star is allowed to change
the value of $N$ or $M$. Such damping mechanisms have not been
considered in our paper. The crucial role of
the migration of unstable soliton ($U$--branch) to a stable soliton
($S$--branch) has also been pointed out in Ref. [28].
According to this paper, a stable soliton ($S$--branch star), which
is slightly perturbed, will oscillate
with a fundamental frequency. This coincides with our conclusion
about the oscillation regime of star near a stable soliton configuration.
Thus the picture of the star's behavior obtained on the basis of
theory of singularities of smooth maps (non--elementary catastrophe
theory) completely coincides with conclusions obtained on the
basis of extensive studies of the numerical evolution of boson
stars.

\bigskip

\centerline{\bf Acknowledgments}
We are grateful to F. W. Hehl for a useful discussion 
and J. A. Wheeler for some important hints.
One of us (F.V.K.) thanks M. L. Ristig for extended 
hospitality at the Institute
for Theoretical Physics, Cologne.

\bigskip

\centerline{\bf REFERENCES}

\newref
[1] J.A. Wheeler, {\it Phys. Rev.} {\bf 97}, 511 (1955).

\newref
[2]  T.D. Lee, {\it Comm. Nucl. Part. Phys.} {\bf 17}, 225 (1987); 
T.D. Lee, {\it Phys. Rev} {\bf D35}, 3637 (1987).

\newref
[3] R. Friedberg, T.D. Lee and Y. Pang, {\it Phys. Rev.}
{\bf D35}, 3640, 3658, 3678 (1987).

\newref
[4]  D.J. Kaup, {\it Phys. Rev.} {\bf 172}, 1331 (1968).

\newref 
[5]  R. Ruffini and S. Bonazzola, {\it Phys. Rev.} {\bf 187}, 1767 (1969).

\newref 
[6]  E.W. Mielke and R. Scherzer, {\it Phys. Rev.} {\bf D24}, 2111 (1981).

\newref 
[7]  Ya.B. Zeldovich and I.D. Novikov:
{\it Stars and Relativity}, Relativistic Astrophysics, Vol.1
(University of Chicago Press, Chicago 1971).

\newref
[8] B.K. Harrison, K.S. Thorne, M. Wakano, J.A. Wheeler:
{\it Gravitation Theory and Gravitational Collapse}
(University of Chicago Press, Chicago 1965).

\newref
[9] S.L. Shapiro and S.A. Teukolsky: {\it Black Holes, White Dwarfs,
and Neutron Stars. The Physics of Compact Objects.}, 
(New York, Wiley 1983).

\newref 
[10] T.D. Lee and Y. Pang , {\it Nucl. Phys.} {\bf B315}, 477 (1989).

\newref 
[11] M. Gleiser and R. Watkins, {\it Nucl. Phys.} {\bf B319}, 733 (1989).

\newref 
[12] Ph. Jetzer, {\it Nucl. Phys.} {\bf B316}, 411 (1989);
{\it Phys. Lett.} {\bf B222}, 447 (1989);
{\it Phys. Lett.} {\bf B243}, 36 (1990).

\newref
[13] M. Gleiser, {\it Phys. Rev.} {\bf D38}, 2376 (1988);
(E) {\it Phys. Rev.} {\bf D39}, 1257 (1989).

\newref
[14] F.V. Kusmartsev, {\it Phys. Rep.} {\bf 183}, 1 (1989).

\newref
[15] V.I. Arnold, {\it Usp. Mat. Nauk.} {\bf 23}, 3 (1968);
{\bf 30}, 3 (1975); V.I. Arnold, S.M. Gusein--Zade, A.N. Varchenko:
{\it Singularities of differentiable maps} (Birkh\"auser, Boston 1985).

\newref
[16] M. Colpi, S.L. Shapiro, and I. Wasserman, {\it Phys. Rev. Lett.} 
{\bf 57}, 2485 (1986).

\newref
[17] F.E. Schunck: {\it Eigenschaften des Bosonen--Sterns}, 
Diploma--thesis, University of Cologne, January 1991.

\newref
[18] P. Baekler, E.W. Mielke, R. Hecht, and F.W. Hehl, {\it Nucl. Phys.}
{\bf 288}, 800 (1987).

\newref
[19]  D. Christodoulou, {\it Commun. Math. Phys.} {\bf 109}, 613 (1987).

\newref
[20] R.C. Tolman, {\it Phys. Rev.} {\bf 35}, 875 (1930); see also
R. Penrose, in: {\it Gravitational Collapse and Relativity}, 
H. Sato and T. Nakamura eds. (World Scientific, Singapure 1986), p.43.

\newref
[21] R. Friedberg, T.D. Lee and A. Sirlin, {\it Phys. Rev.} {\bf D13},
2739 (1976).

\newref
[22] T. Poston, I. Stewart: {\it Catastrophe Theory and its 
Applications} (Pitman, 1978).

\newref
[23] F.V. Kusmartsev, {\it Phys. Rev.} {\bf B43}, 1345 (1991).

\newref
[24] F.V. Kusmartsev and E.I. Rashba, {\it Sov. Phys. --- JETP}
{\bf 57}, 1202 (1983).

\newref
[25] F.V. Kusmartsev, {\it Physica Scripta} {\bf 29}, 7 (1984).

\newref
[26] S. Bahcall, B.W. Lynn, and S. Selipsky, {\it Nucl. Phys.} 
{\bf B325}, 606 (1989); {\bf 331}, 67 (1990).

\newref
[27] B. Gradwohl and G. K\"albermann, {\it Nucl. Phys.} {\bf B324},
215 (1989).

\newref
[28] E. Seidel and W.--M. Suen, {\it Phys. Rev.} {\bf D42}, 384 (1990).

\pagebreak

\begin{figure}[ht]
\centering 
\leavevmode\epsfysize=20cm
\epsfbox{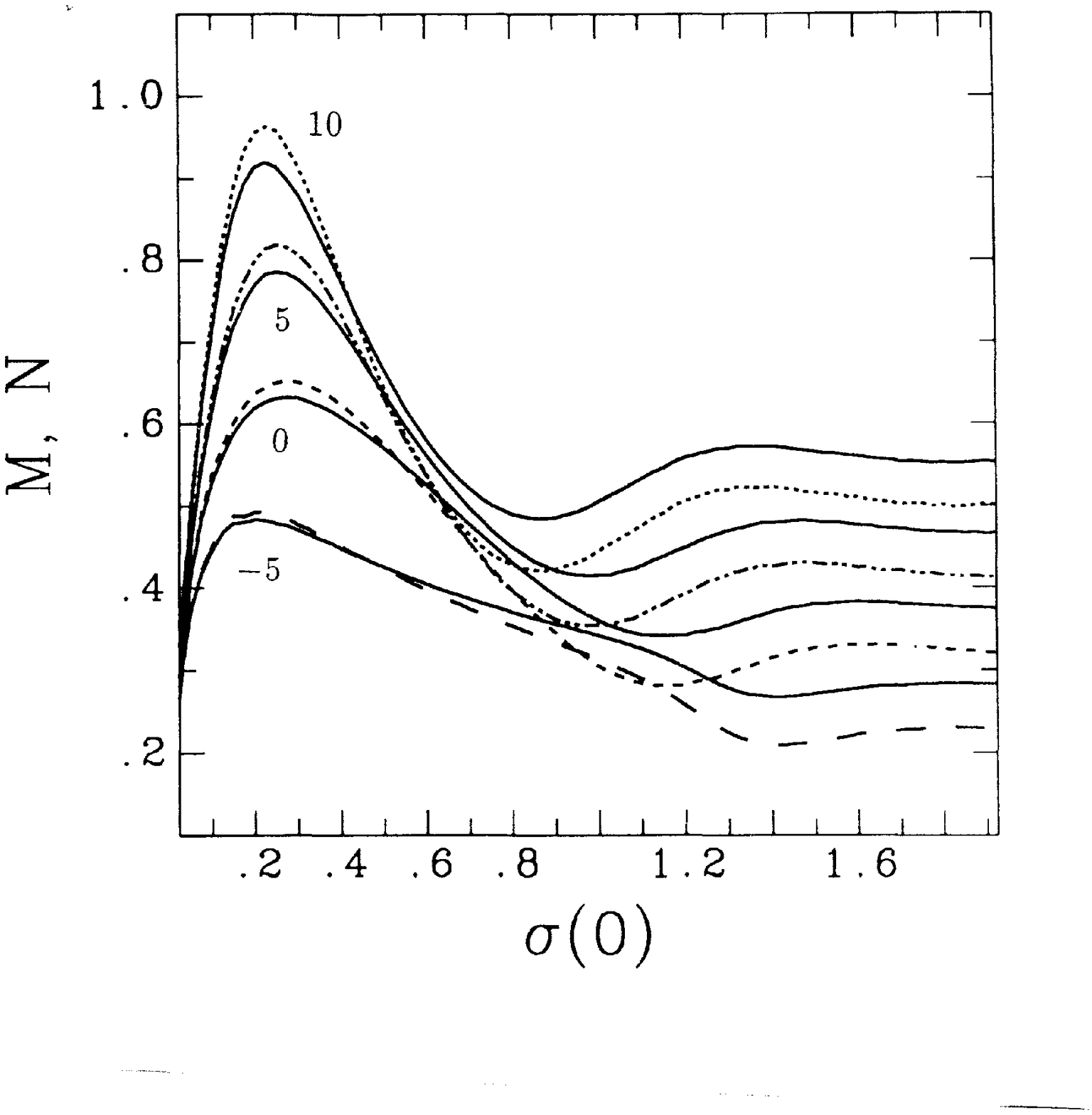}\\ \vskip 1cm
\caption[fig1]{The Tolman mass $M$ (---) in units of ($1/mG$)
and particle number $N$ ($-$ $-$) in dimensionless units of ($1/m^2G$)
as a function of the central density 
$\sigma (0)=\sqrt{\kappa /2}\mid \Phi (0)\mid$ for various
$\tilde \alpha :=(2 \alpha /\kappa m^2)$ and $\beta =0$
in the potential $U$. For a linear scalar field 
($\tilde \alpha =0$) the Kaup limit $M_{Kaup}=0.633$ is recovered [17].
The maxima and minima correspond to the $A_2$ singularities.}
\end{figure}

\begin{figure}[ht]
\centering 
\leavevmode\epsfysize=20cm
\epsfbox{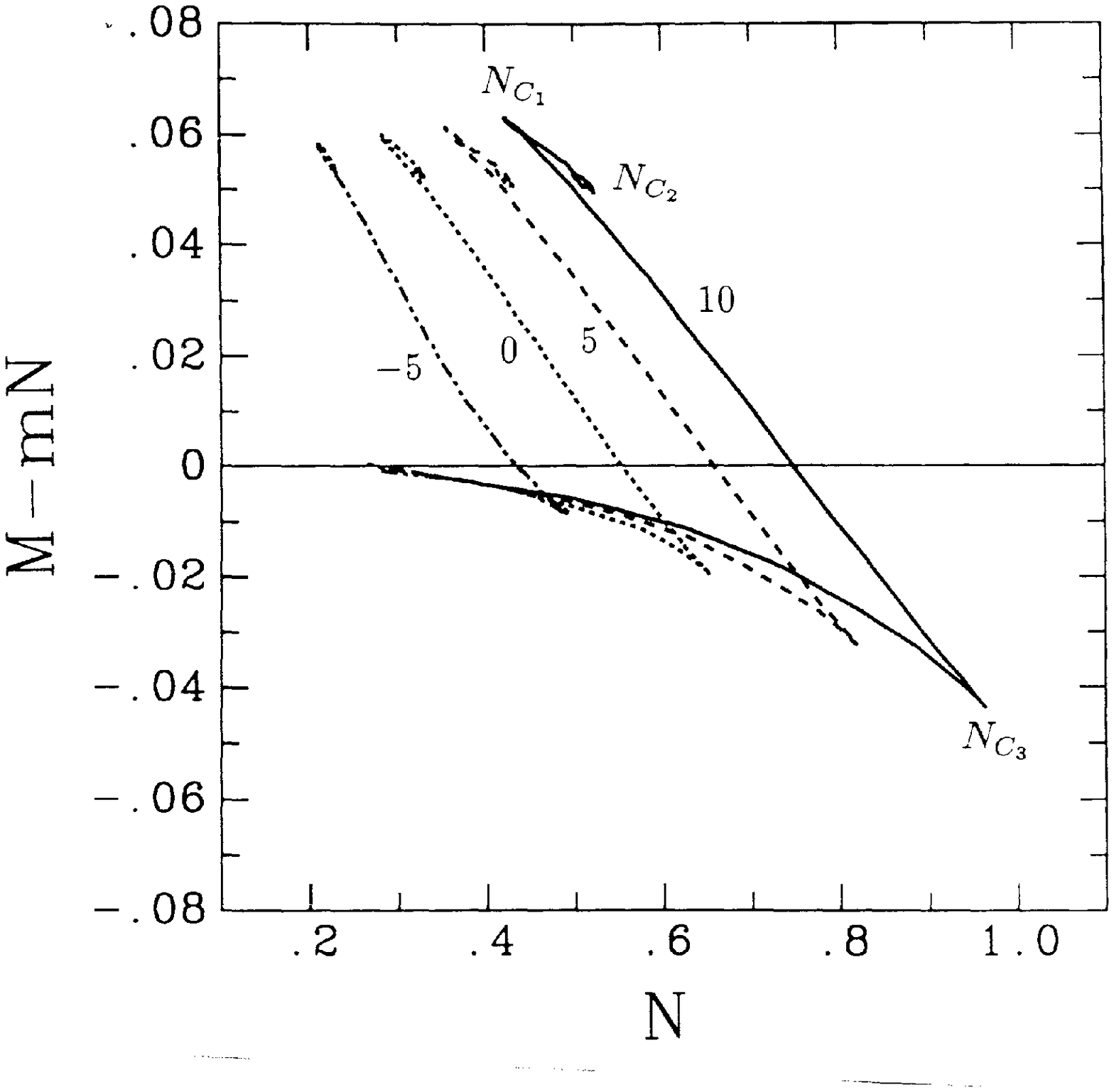}\\ \vskip 1cm
\caption[fig2]{The binding energy $M-mN$ as a function of $N$ at 
different values $\tilde \alpha =-5,0,5,10$ [17].
In this bifurcation diagram, 
the lower branch of each cusp corresponds 
to a stable star configuration. See, for comparison, Fig. 1.}
\end{figure}

\begin{figure}[ht]
\centering 
\leavevmode\epsfysize=20cm
\epsfbox{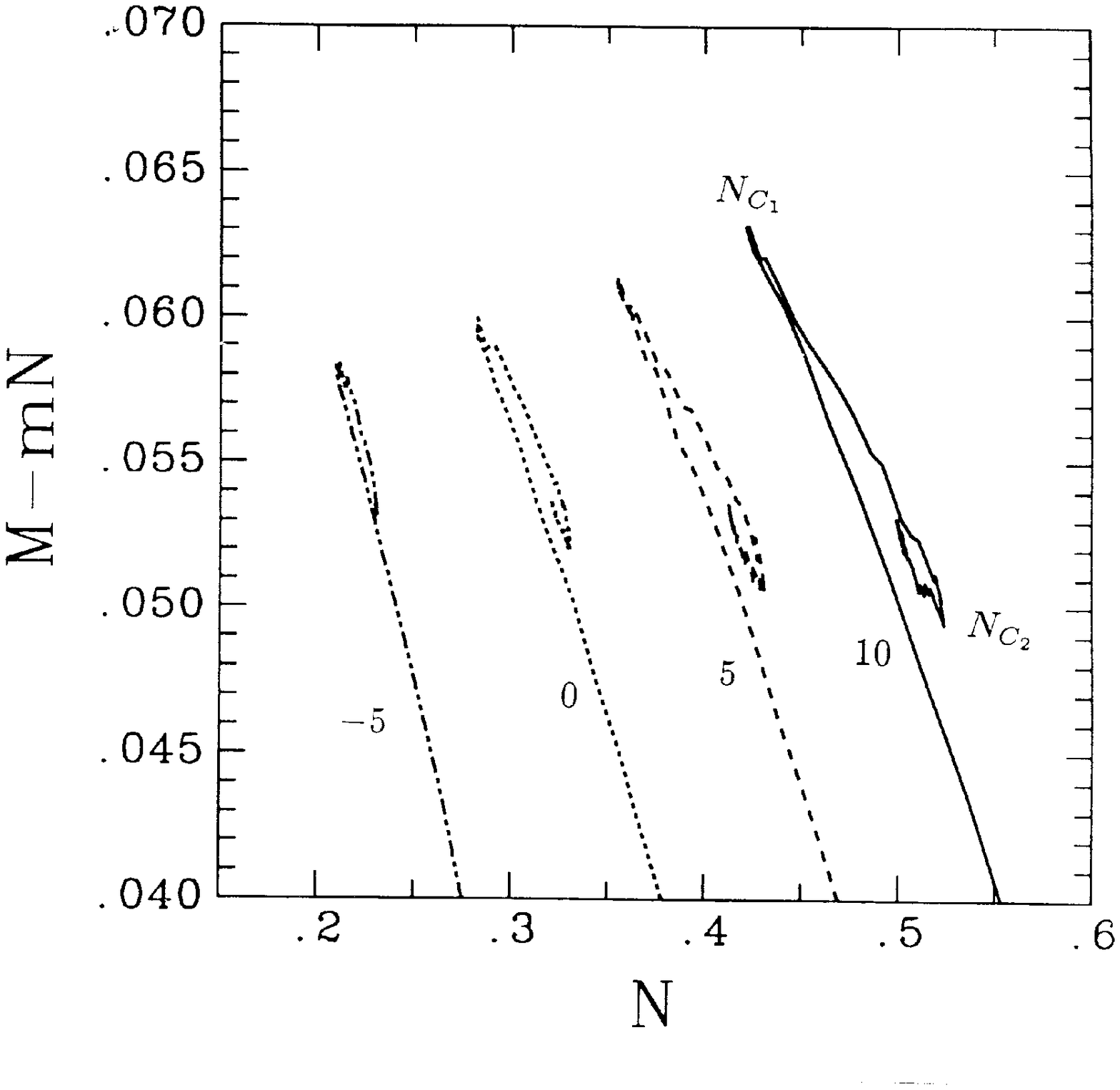}\\ \vskip 1cm
\caption[fig3]{``Magnified" view of the binding energy for the same
parameters as in Fig. 2.}
\end{figure}

\begin{figure}[ht]
\centering 
\leavevmode\epsfysize=20cm
\epsfbox{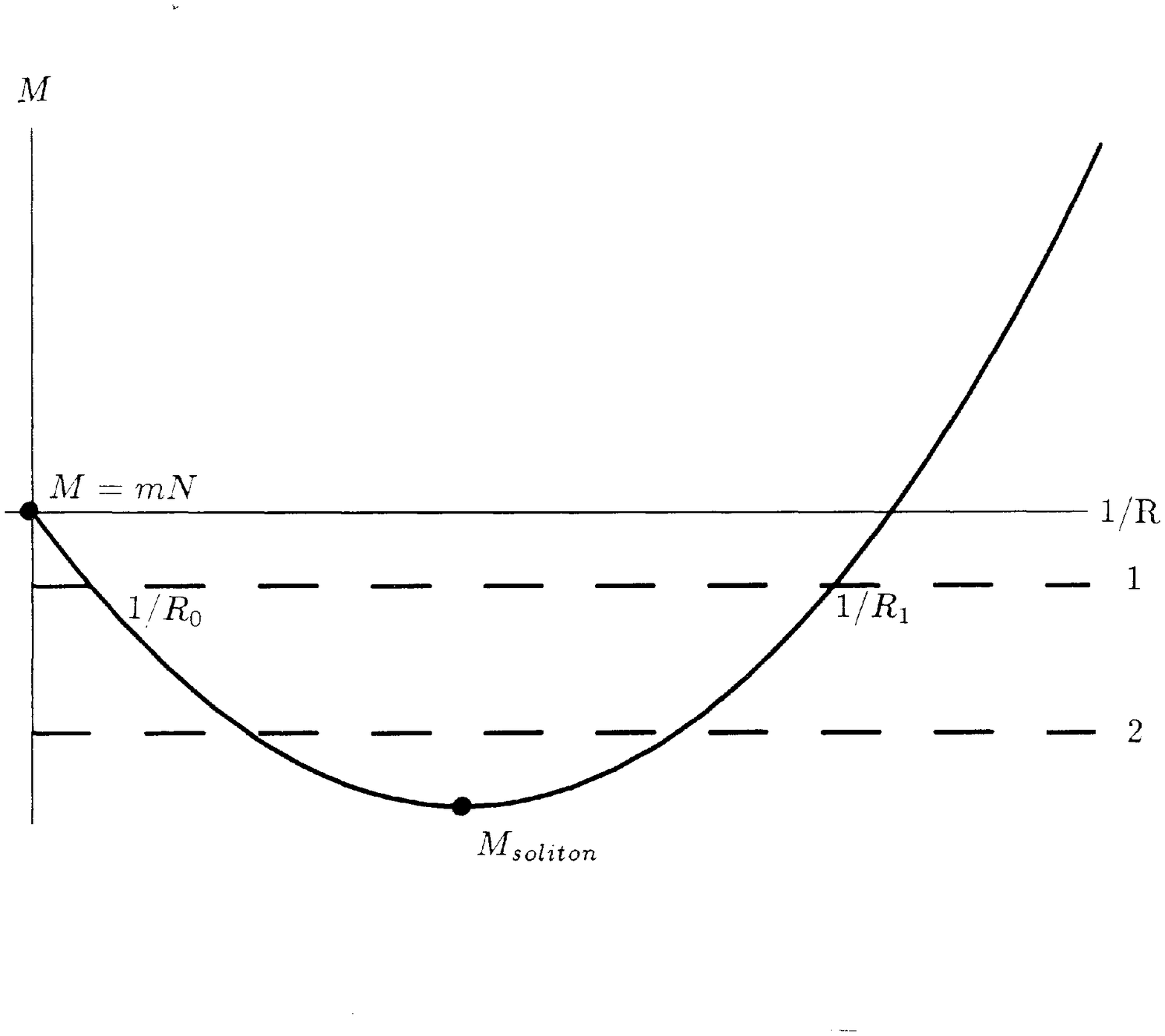}\\ \vskip 1cm
\caption[fig4]{``Adiabatic'' potential $M=M(1/R)$
of the star for $N<N_{C_1}$
(Schematic construction following Ref. $[22, 24]$).}
\end{figure}

\begin{figure}[ht]
\centering 
\leavevmode\epsfysize=20cm
\epsfbox{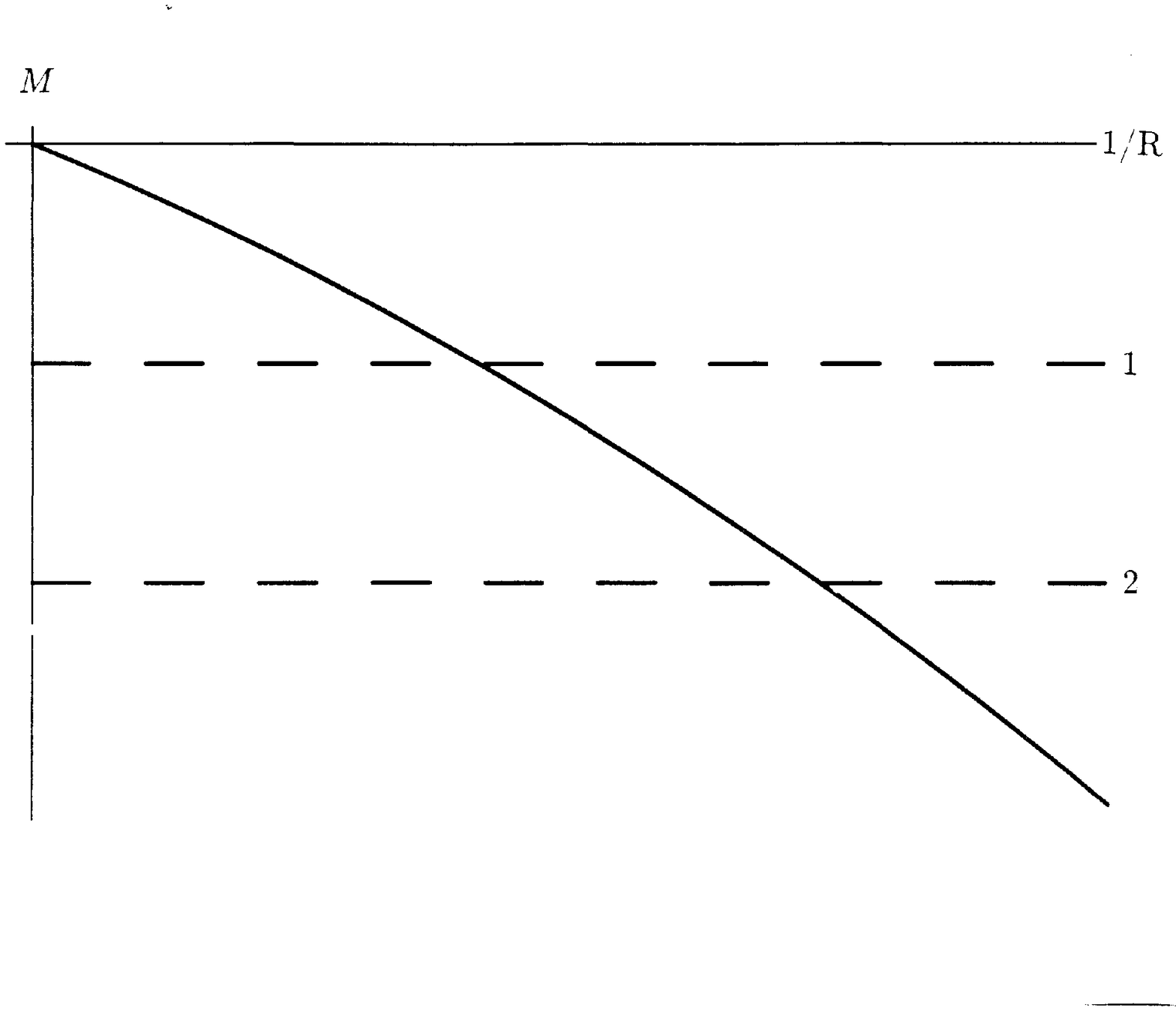}\\ \vskip 1cm
\caption[fig5]{``Adiabatic'' potential $M=M(1/R)$ of the star
for $N>N_{C_3}$, describing the collapse.}
\end{figure}

\begin{figure}[ht]
\centering 
\leavevmode\epsfysize=20cm
\epsfbox{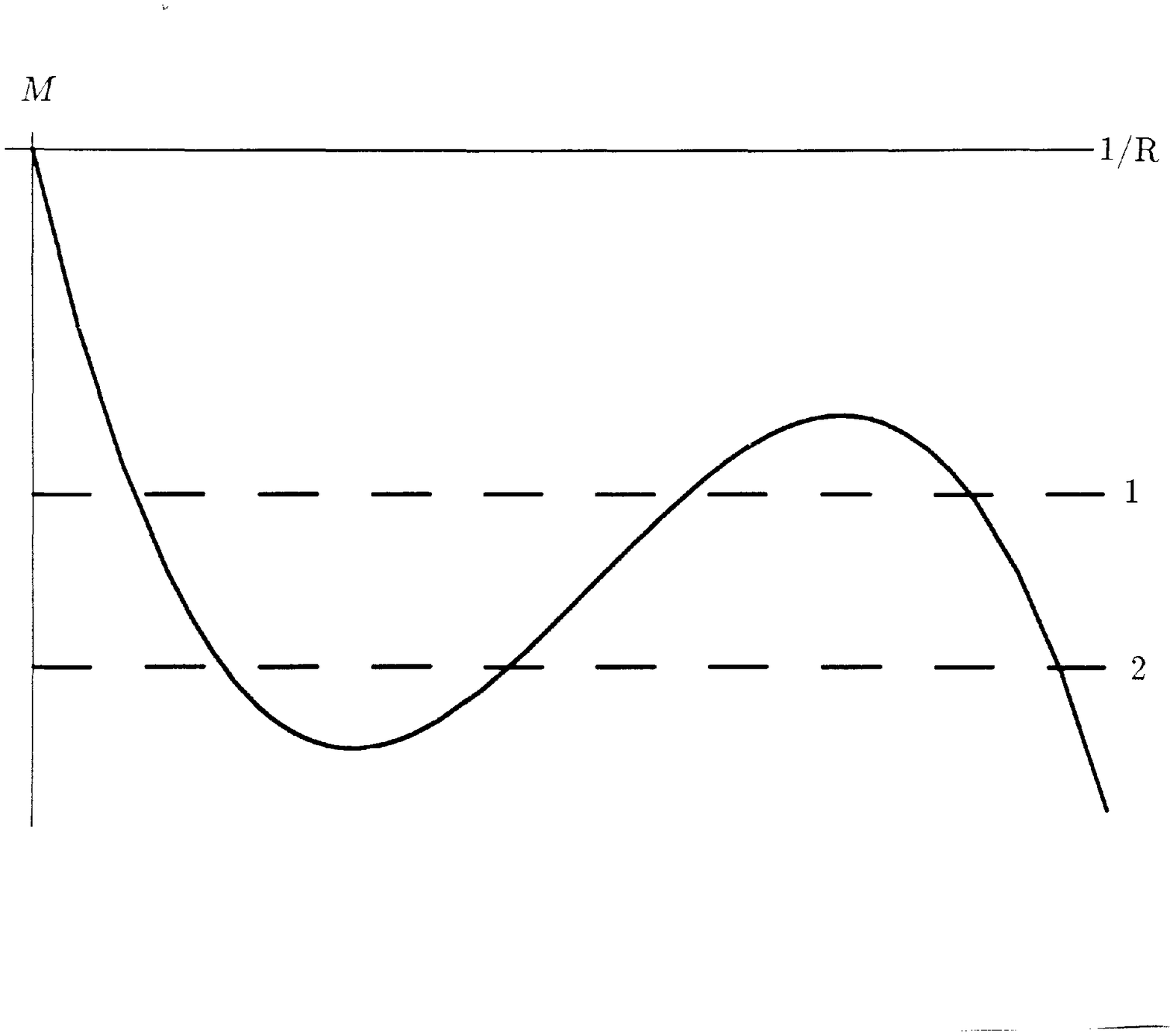}\\ \vskip 1cm
\caption[fig6]{``Adiabatic'' potential $M=M(1/R)$ of the star for 
$N\in [N_{C_2}, N_{C_3}] $, describing oscillation and collapse.}
\end{figure}

\end{document}